\documentclass[prd,tightenlines,nofootinbib,superscriptaddress,showpacs]{revtex4}
\usepackage[utf8x]{inputenc}
\usepackage{amsmath} 
\usepackage{amsfonts,hyperref}
\usepackage{slashed}
\usepackage[dvips]{graphicx}
\usepackage{color}
\usepackage{units}

\allowdisplaybreaks

\newcommand{\und}{{\text{ and }}}

\newcommand{\Rect}{{\text{Rect}}}
\newcommand{\sign}{{\text{sign}}}
\newcommand{\E}{{\text{\textbf{E}}}}
\newcommand{\K}{{\text{\textbf{K}}}}
\newcommand{\PiI}{{\boldsymbol{\Pi}}}

\newcommand{\cv}{{c}}

\renewcommand{\Im}{{\text{Im}}}
\renewcommand{\Re}{{\text{Re}}}

\newcommand{\e}{{\text{e}}}
\renewcommand{\i}{{\text{i}}}
\renewcommand{\e}{{\text{e}}}

\newcommand{\scE}{{\mathcal{E}}}

\newcommand{\Eq}{{Eq.~}}
\newcommand{\Eqs}{{Eqs.~}}
\newcommand{\varpm}{-}
\newcommand{\varmp}{+}
\newcommand{\mppm}{{\pm}}
\newcommand{\pmmp}{{\pm}}

\begin{document}
 \author{{\bf Eckhard Strobel}\email{eckhard.strobel@gravity.fau.de}}
 \affiliation{ICRANet, Piazzale della Repubblica 10, 65122 Pescara, Italy}
 \affiliation{Dipartimento di Fisica, Universit\`a di Roma "La Sapienza", Piazzale Aldo Moro 5, 00185 Rome, Italy}
 \affiliation{Universit\'e de Nice Sophia Antipolis, 28 Avenue de Valrose, 06103 Nice Cedex 2, France}
 \author{{\bf She-Sheng Xue}\email{xue@icra.it}}
 \affiliation{ICRANet, Piazzale della Repubblica 10, 65122 Pescara, Italy}
 \affiliation{Dipartimento di Fisica, Universit\`a di Roma "La Sapienza", Piazzale Aldo Moro 5, 00185 Rome, Italy}

\date{\today}

\title{Semiclassical pair production rate for rotating electric fields}


\begin{abstract}
We semiclassically investigate Schwinger pair production for pulsed rotating electric fields depending on time. To do so we solve the Dirac equation for two-component fields in a WKB-like approximation. The result shows that for two-component fields the spin distribution of produced pairs is generally not \(1:1\). As a result the pair creation rates of spinor and scalar quantum electro dynamics (QED) are different even for one pair of turning points. For rotating electric fields the pair creation rate is dominated by particles with a specific spin depending on the sense of rotation for a certain range of pulse lengths and frequencies. We present an analytical solution for the momentum spectrum of the constant rotating field. We find interference effects not only in the momentum spectrum but also in the total particle number of rotating electric fields.
\end{abstract}

\maketitle

\section*{Introduction}
Since the first investigations of electron-positron pair creation in strong electric fields, also known as the Schwinger effect, there has been a lot of theoretical investigations of it. However it has not yet been possible to measure it directly due to the exponentially damped pair creation rate \(\sim\exp(-\pi /\epsilon)\)  where \(\epsilon=E/E_c\) is the field strength \(E\) normalized by the critical electric field \cite{Sauter1931,Heisenberg1936,Schwinger1951} 
\begin{align}
 E_c=\frac{m^2c^3}{e\hbar}.
\end{align}
As laser powers in future may get closer to reaching this critical field strength, investigations of the effect are of interest. However there is the possibility that different strong field processes, such as QED cascades, will set in which might prevent reaching critical intensities \cite{Bell2008,Kirk2009,Fedotov2010,Bulanov2010B,Elkina2011,Nerush2011,Nerush2011B,King2013,Bulanov2013,Gonoskov2013,Bashmakov2014}.\\ 
As pointed out in \cite{Blinne2013} it might be crucial for the detection of the Schwinger effect to evaluate if one can distinguish a QED cascade triggered by an electron which was produced in the Schwinger process from one which was triggered by vacuum impurities. To do so one first has to evaluate the properties of the pairs produced via the Schwinger effect. The simplest field configurations taken into account for cascade calculations are uniformly rotating electrical fields. These are approximative models for the electric fields in the anti-nodes of circularly polarized standing waves. \\
The momentum spectrum of produced electron-positron pairs for rotating fields has been recently investigated numerically for pulsed fields with the help of the real-time Dirac-Heisenberg-Wigner (DHW) formalism \cite{Blinne2013} as well as analytically for constant rotating fields with help of the semiclassical WKB approximation for spinor QED \cite{paper1}. Numerical methods  usually need a lot of computation time. Semiclassical methods can help to give a better understanding of special features of the momentum spectrum, e.g. interference effects.\\
For one-component fields depending solely on time there exist a lot of semiclassical investigations using e.g. the WKB-approximation \cite{Brezin1970,Popov1971,Popov1972,Popov1973,Marinov1977,Popov2001,Kleinert2008,Dumlu2010,Dumlu2011,Kleinert2013} or the world-line instanton method \cite{Dunne2005B,Dunne2006,Dumlu2011WLI}. In addition to that there are exact solutions for fields depending on lightcone variables \cite{Tomaras2001,Hebenstreit2011B}. While one-component fields can be used to describe linearly polarized laser fields, one needs two-component fields to study pair creation of circularly polarized fields. There exist some semiclassical studies of more involved fields such as electric fields with two or three components depending on space \cite{Dunne2006B} or fields depending on space and on time \cite{Schneider2014}. The exponential factor of rotating electric fields was studied in \cite{Popov1973} with the imaginary time method and in \cite{Xie2012} with the world-line instanton method.\\
In \cite{paper1} the WKB method was used to compute the momentum spectra for more than one-component fields in scalar QED. The assumption was made that for one pair of turning points the results of spinor and scalar QED are the same at the leading non perturbative order.  This was based on the fact that it was shown to be true for one-component fields in \cite{Dumlu2011}.\\
In the present work we show that for two-component fields there is a difference in the pair creation rate of spinor and scalar QED even if there is only one pair of turning points. To do so we start from the Dirac equation and find a semiclassical approximation which is a generalization of the WKB-approximation for the one-component case. We find that in the two-component case there is a factor depending on the spin of the produced pairs compared to the scalar case. This factor gives rise to  a different rate for particles in different spin states.\\
Looking at examples of rotating electric fields we find that particles of one spin direction dominate the produced pairs depending on the sense of rotation of the field for a certain range of pulse lengths and frequencies. For the constant rotating field a generalization of the analytic results for the scalar case \cite{paper1} is found, so that the momentum spectrum can be calculated analytically.\\ 
This work is arranged as follows: In Section \ref{sec:SCMomSpec} we derive the semiclassical momentum spectrum for two-component electric fields depending on time. These results are used in Section \ref{sec:rotating}  
to study the momentum spectrum of the constant rotating field.
The total pair creation rate of this field is investigated in Section \ref{sec:TotalParticleYield}. Section \ref{sec:conclusions} contains our conclusions and remarks. In order to keep the main text and ideas more clear, we have relegated some of the technical calculations to the Appendix. In Appendix \ref{app:Ansatz} we motivate our ansatz for the coefficients used in Section \ref{sec:SCMomSpec}. The analytic calculations necessary for the computation of the momentum spectra of the constant rotating field 
can be found in Appendix \ref{app:RotAna}.

\section{Semiclassical momentum spectrum}
\label{sec:SCMomSpec}
Here we compute the semi classical momentum spectrum for spinor QED for fields with two time-dependent components. The corresponding scalar case has been treated in \cite{paper1}. We start from the Dirac equation
\begin{align}
 \left(\left[\i \hbar \partial_\mu-e A_\mu(x)\right]\gamma^\mu-mc\right)\Psi(\vec{x},t)=0.
\end{align}
We can go to the squared version by substituting
\begin{align}
 \Psi(\vec{x},t)=\left[\left(\i \hbar \partial_\mu-e A_\mu(x)\right)\gamma^\mu+mc\right]\psi(\vec{x},t),
\end{align}
for which we find
\begin{align}
 \left[\left(\i \hbar \partial_\mu-e A_\mu(x)\right)^2-\hbar\frac{e}{2}F_{\mu\nu}(x)\sigma^{\mu\nu}-m^2c^2\right]\psi(\vec{x},t)=0,
\end{align}
where 
\begin{align}
 \sigma^{\mu\nu}=\frac{\i}{2}[\gamma^\mu,\gamma^\nu]
\end{align}
and
\begin{align}
 F_{\mu\nu}(x)=\partial_\mu A_{\nu}(x)-\partial_\mu A_\nu(x).
\end{align}
For solely time-dependent fields with the general potential
\begin{align}
 A_\mu(x)=\frac{1}{e c}\left(0,\vec{V}(t)\right)
\end{align}
the field strength tensor only has non vanishing components for \(\mu=0\) or \(\nu=0\). \\
We now choose to work in the Weyl basis, i.e.
\begin{align}
 \gamma^{j}=\begin{pmatrix}
             0 &\sigma^j\\
             -\sigma^j & 0
            \end{pmatrix},
&&
  \gamma^{0}=\begin{pmatrix}
             0 &I_2\\
             I_2 & 0
            \end{pmatrix},
&& \text{ where }&&
   \sigma^x=\begin{pmatrix}
             0 & 1\\
             1 & 0
            \end{pmatrix},
&&
   \sigma^y=\begin{pmatrix}
             0 & -\i\\
             \i & 0
            \end{pmatrix},
            &&
   \sigma^z=\begin{pmatrix}
             1 & 0\\
             0 & -1
            \end{pmatrix}.
\end{align}
In this basis we find
\begin{align}
 \sigma^{0j}=\i\begin{pmatrix}
             -\sigma^j & 0\\
             0 & \sigma^j
            \end{pmatrix}.
\end{align}
If we now use the ansatz
\begin{align}
 \psi(\vec{x},t)=\begin{pmatrix}
          \psi_+(t)\\
          \psi_-(t)
         \end{pmatrix}
\e^{-\i\frac{\vec{P}\cdot\vec{x}}{\hbar}},
&& \text{where} &&
\psi_\pm(t)=\begin{pmatrix}
            \psi_1^\pm(t)\\
            \psi_2^\pm(t) 
            \end{pmatrix}
\end{align}
and \(\vec{P}\) is the canonical momentum, we find the four coupled equations
\begin{align}
 \left[\hbar^2\partial_t^2+\scE(t)^2\mp\i\hbar\, c\dot{p}_z(t)\right]\psi_1^\pm(t)\mp\i\hbar\, c\dot{p}_{x-y}(t)\,\psi_2^\pm(t)&=0, \label{eq:coupledDirac1}\\
 \left[\hbar^2\partial_t^2+\scE(t)^2\pm\i\hbar\, c\dot{p}_z(t)\right]\psi_2^\pm(t)\mp\i\hbar\, c\dot{p}_{x+y}(t)\,\psi_1^\pm(t)&=0,
 \label{eq:coupledDirac2}
\end{align}
where we defined
\begin{align}
\scE(t)^2:=c^2\vec{p}(t)^2+m^2c^4, \label{eq:scE} \\
p_{x\pm y}(t):=p_x(t)\pm\i p_y(t) 
\end{align}
and introduced the kinetical momentum 
\begin{align}
 c\vec{p}(t):=c\vec{P}-\vec{V}(t).
\end{align}
Because of the chiral properties of the Weyl-Basis we can interpret \(\psi^+(t)\) and \(\psi^-(t)\) as different spin states of the produced pairs.\\
For the two-component case \(V_z(t)=0\) we can now make the ansatz (for a motivation see Appendix \ref{app:Ansatz})
\begin{align}
 \psi^{s}_{1}(t)&=\frac{\sqrt{cp_{x-y}(t)}}{\sqrt{\scE(t)}}\sqrt{cp_\parallel(t)}C^s\left(\alpha^s(t)\frac{\e^{-\frac{\i}{2}K_s(t)}}{\sqrt{\scE(t)-s\epsilon_\perp}} +\beta^s(t)\i\frac{\e^{\frac{\i}{2}K_s(t)}}{\sqrt{\scE(t)+s\epsilon_\perp}}\right),\label{eq:ansatz1} \\
 \dot{\psi}^{s}_{1}(t)&=-\i\frac{\scE(t)}{\hbar}\frac{\sqrt{cp_{x-y}(t)}}{\sqrt{\scE(t)}}\sqrt{cp_\parallel(t)}C^s\left(\alpha^s(t)\frac{\e^{-\frac{\i}{2}K_s(t)}}{\sqrt{\scE(t)-s\epsilon_\perp}} -\beta^s(t)\i\frac{\e^{\frac{\i}{2}K_s(t)}}{\sqrt{\scE(t)+s\epsilon_\perp}}\right) ,\label{eq:ansatzdot1} \\
  \psi^{s}_{2}(t)&=-s\frac{\sqrt{cp_{x+y}(t)}}{\sqrt{\scE(t)}}\sqrt{cp_\parallel(t)}C^s\left(\alpha^s(t)\frac{\e^{-\frac{\i}{2}K_s(t)}}{\sqrt{\scE(t)+s\epsilon_\perp}} -\beta^s(t)\i\frac{\e^{\frac{\i}{2}K_s(t)}}{\sqrt{\scE(t)-s\epsilon_\perp}}\right), \label{eq:ansatz2} \\
 \dot{\psi}^{s}_{2}(t)&=s\i\frac{\scE(t)}{\hbar}\frac{\sqrt{cp_{x+y}(t)}}{\sqrt{\scE(t)}}\sqrt{cp_\parallel(t)}C^s\left(\alpha^s(t)\frac{\e^{-\frac{\i}{2}K_s(t)}}{\sqrt{\scE(t)+s\epsilon_\perp}} +\beta^s(t)\i\frac{\e^{\frac{\i}{2}K_s(t)}}{\sqrt{\scE(t)-s\epsilon_\perp}}\right), \label{eq:ansatzdot2} 
\end{align}
where we defined the spin parameter \(s:=\pm1\) and
 \begin{align}
 \epsilon_\perp^2&:=c^2P_z^2+m^2c^4,\\
 p_\parallel(t)^2&:=p_x(t)^2+p_y(t)^2,
 \end{align}
 as well as the integrals
 \begin{align}
 K_\pm(t)&:=K(t)\pm K_{xy}(t),\label{eq:K_s}\\
K(t)&:=\frac{2}{\hbar}\int_{0}^{t} \scE(t') dt', \label{eq:Kint}\\
 K_{xy}(t)&:=\epsilon_\perp\int_{0}^{t} \frac{\dot{p}_x(t')p_y(t')-\dot{p}_y(t')p_x(t')}{\scE(t')p_\parallel(t')^2}dt' \label{eq:K_xy}.
 \end{align}
Using the ansatz (\ref{eq:ansatz1})-(\ref{eq:ansatzdot2}) in the Dirac equation (\ref{eq:coupledDirac1})-(\ref{eq:coupledDirac2}) leads to coupled equations for the coefficients
\begin{align}
 \dot{\alpha}^s(t)&=\frac{\dot{\scE}(t)}{2 \scE(t)}G_+(t) \e^{\i K_{s}(t)}\beta^s(t),\label{eq:alpha}\\
 \dot{\beta}^s(t)&=\frac{\dot{\scE}(t)}{2 \scE(t)}G_-(t) \e^{-\i K_{s}(t)}\alpha^s(t),\label{eq:beta}
\end{align}
where
\begin{align}
 G_\pm(t)=-s\frac{\i\epsilon_\perp}{ cp_\parallel(t)}\pm \frac{\dot{p}_x(t)p_y(t)-\dot{p}_y(t)p_x(t)}{\dot{p}_x(t)p_x(t)+\dot{p}_y(t)p_y(t)}\frac{\scE(t)}{ cp_\parallel(t)}.
\end{align}
As shown in Appendix \ref{app:Ansatz} this solution reduces correctly to the known results of the one-component case for \(A_y(t)=0\).\\
The number of produced electron-positron pairs as a function of the momentum \(\vec{P}\) can be found  as the transmission probability
\begin{align}
 W^s(\vec{P}):=\lim_{t\rightarrow\infty} \left|\beta^s(t)\right|^2. \label{eq:trans}
\end{align}
If we now use appropriate boundary conditions \cite{Dumlu2011}
\begin{align}
 \beta^s(-\infty)=0, && \alpha^s(-\infty)=1, 
\end{align}
 we can by iteratively using \Eqs (\ref{eq:alpha}) and (\ref{eq:beta})  find
 \begin{align}
 \beta^s(\infty)=\sum_{m=0}^\infty \int_{-\infty}^\infty dt_0G_-(t_0)\frac{\dot{\scE}(t_0)}{2\scE(t_0)}\e^{-\i K_s(t_0)}\prod_{n=1}^m \int_{-\infty}^{t_{n-1}} d\tau_nG_+(\tau_n)\frac{\dot{\scE}(\tau_n)}{2\scE(\tau_n)}\e^{\i K_{s}(\tau_n)} \int_{-\infty}^{\tau_n}dt_nG_-(t_n)\frac{\dot{\scE}(t_n)}{2\scE(t_n)}\e^{-\i K_s(t_n)}.\label{eq:multiint}
\end{align}
Note that this was done by analogy with \cite{Berry1982}. These integrals are dominated by the regions around the turning points which are given by
\begin{align}
  \scE(t_p^{\pm}):=0 \label{eq:turningpoints}.
\end{align} 
According to \Eq(\ref{eq:scE}) the \(t_p^{\pm}\) are found in complex conjugated pairs. Following from \Eq(\ref{eq:multiint}) by deforming the contour we extract the singularities for the turning points \(t_p:=t_p^+\) lying in the upper half plane of the imaginary plane.  By analogy with \cite{Berry1982} we consider general singularities of order \(\nu\)
\begin{align}
 \scE(t)\approx A (t-t_p)^{\nu/2},
\end{align}
which leads to
\begin{align}
  K(t)&\approx K(t_p)+\frac{2}{\hbar}\frac{2}{\nu+2}A(t-t_p)^{\nu/2+1},\\
  \frac{\dot{\scE(t)}}{\scE(t)}&\approx\frac{dK(t)}{dt}\frac{\nu}{\nu+2}\frac{1}{K(t)-K(t_p)}.
\end{align}
Using the fact that \(G_\pm(t_p)=s\)  one can now change variables to \(\xi_n=K(t_p)-K(t_n) \und \eta_n=K(t_p)-K(\tau_n)\) to find
\begin{align}
 \beta^s(\infty)\approx-s \sum_{t_p}^\infty  2\pi \i\, \e^{-\i K_s(t_p)}\sum_{m=0}^\infty\left(\frac{\nu_{t_p}}{2(\nu_{t_p}+2)}\right)^{m+1}(-1)^m I_m, \label{eq:solmulti}
\end{align}
where \cite{Berry1982}
\begin{align}
 I_m=\frac{1}{2\pi\i}\int_{-\infty}^\infty d\xi_0\frac{\e^{\i \xi_0}}{\xi_0}\prod_{n=1}^m \int_{\infty}^{\xi_{n-1}} d\eta_n\frac{\e^{-\i \eta_n}}{\eta_n} \int_{\eta_n}^{\infty}d\xi_n\frac{\e^{\i \xi_n}}{\xi_n}=\frac{\pi^{2m}}{(2m+1)!}.
\end{align}
This leads to
\begin{align}
\beta^s(\infty)\approx-2\sum_{t_p}\e^{-\i K_s(t_p)} \sin\left(\frac{\pi\nu_{t_p}}{2(\nu_{t_p}+2)}\right). \label{eq:scalarbinf}
\end{align}
For all the examples covered in this work we deal with simple turning points, i.e.~\(\nu=1\). Thus the momentum spectrum of the pair creation rate (\ref{eq:trans}) in the semiclassical approximation takes the form  
\begin{align}
 W^s_{SC}=\left|\sum_{t_p}\e^{-\i K_s(t_p)}\right|^2 \label{eq:MomentumSpectrum}.
\end{align}
Observe that unlike in the case of one-component fields the number of pairs produced in both spin states is not the same. In fact looking at examples we will see that the factor  \(\exp(\pm\i K_{xy}(t_p))\) usually elevates the pair creation rate for one spin and suppresses the other. \\
Comparing to the one-component case of \cite{Dumlu2011} we find that the factor \((-1)^p\) is replaced by the factor \(\exp(\pm\i K_{xy}(t_p))\) in the two-component case. Accordingly we find from \Eq(\ref{eq:comptoonecomp})
\begin{align}
 \left.e^{\pm \i K_{xy}(t_p)}\right|_{p_y(t)=P_y}=\pm\i\,\sign\left(cp_x(t_p)\right). 
\end{align}
It is also noteworthy that the case of scalar QED which was treated in \cite{paper1} for general time dependent fields is included in this result if one sets \(s=0\).

\section{Momentum Spectrum of constant rotating fields}
\label{sec:example}
\label{sec:rotating} 
\begin{figure}
\centering
 \begin{minipage}{0.5\textwidth}
  \includegraphics[width=1\textwidth]{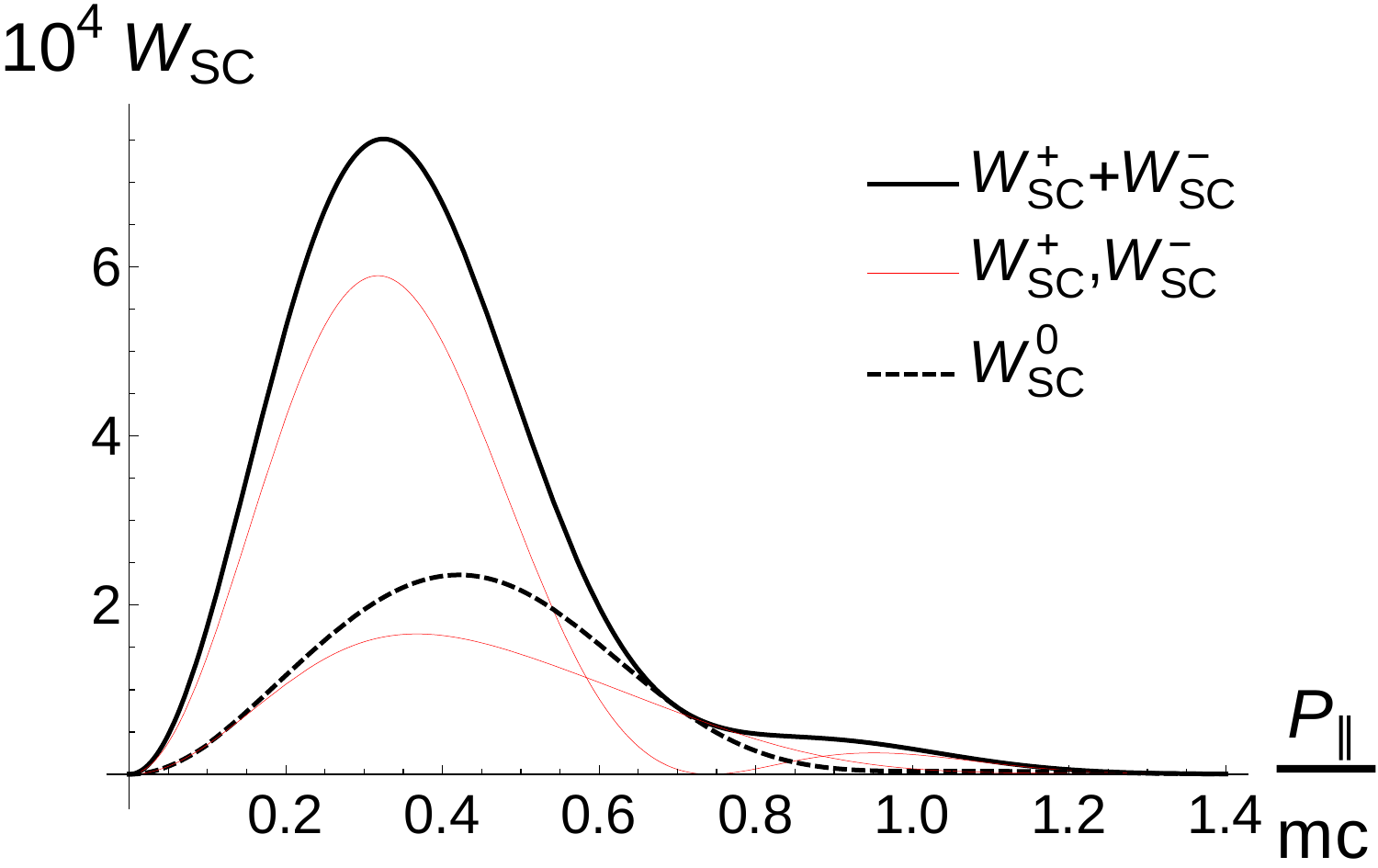}
 \end{minipage}
\begin{minipage}{0.49\textwidth}
  \includegraphics[width=1\textwidth]{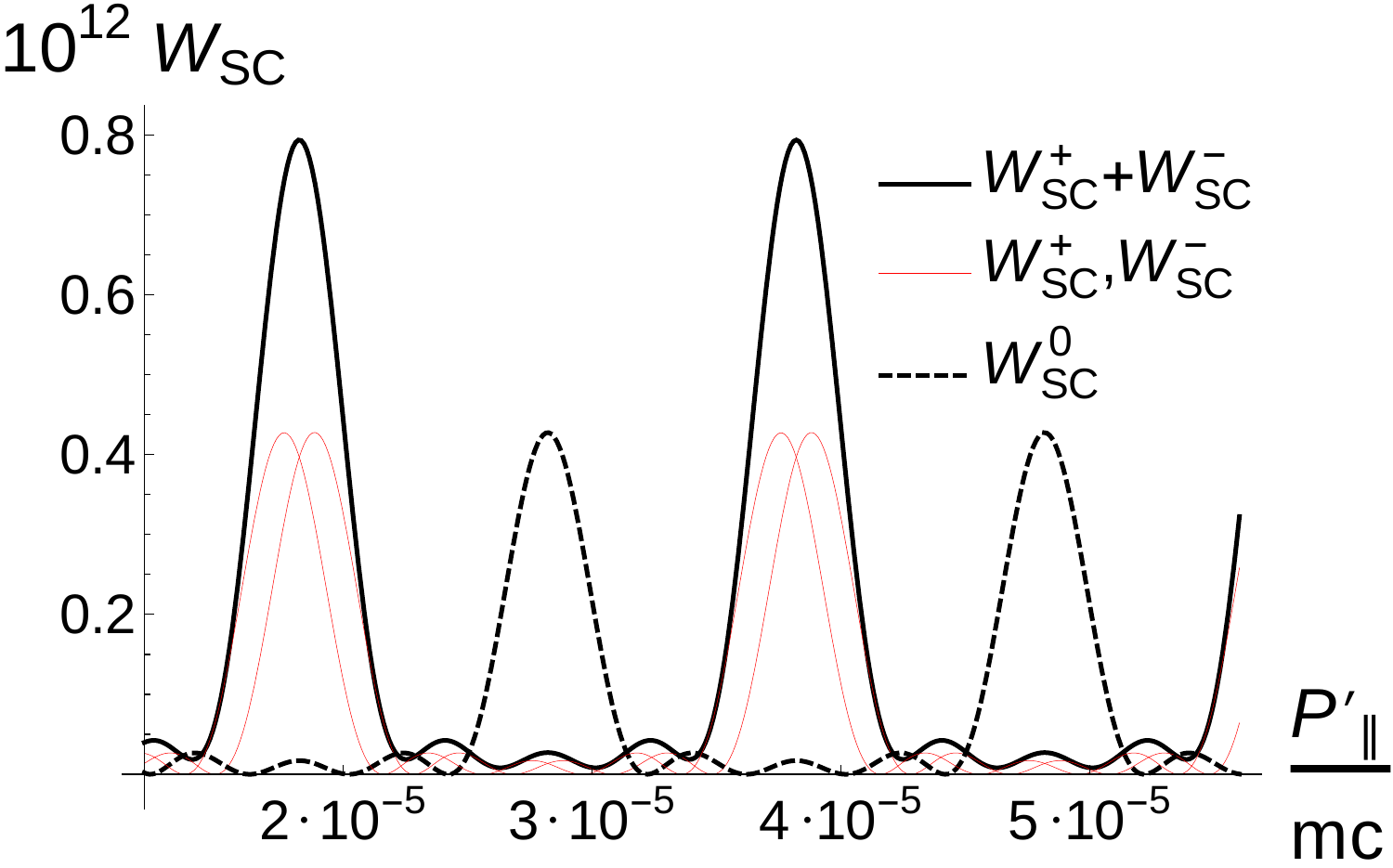}
 \end{minipage}
  \caption{Momentum spectrum of pairs produced by the rotating electric field pulse given in (\ref{eq:constrectpulse}) for an amplitude field strength \(\epsilon=0.1\) and transversal momentum \(P_z=0\) in scalar QED (dashed) as well as in spinor QED (solid). The two different spin states are plotted as a thin red line. The left side shows the spectrum for \( \sigma= 4\pi\,\) and a pulse length of \(\tau=4\pi \lambda_c/c\), which are parameters comparable to the ones of \cite{Blinne2013}. One sees that one spin state depending on the sense of rotation \(g\) dominates the spectrum. On the right sight one can see a part of the spectrum for \(\tau=10\,\text{fs}\) and \(\sigma=10\pi\) (which corresponds to a frequency of \(\nu=500\,\text{THz}\)) as a function of the displaced radial momentum \(P'_\parallel:=P_\parallel-3933.1\,mc\). For these parameters the spectra of the two spin states are merely displaced with respect to the scalar one and each other. Observe the scale of the plot which lets conclude that interference effects will be smeared out by a detector. A plot omitting these effects for a wider range of momenta can be found in Fig.~\ref{fig:T}. } 
  \label{fig:MS_t=infty}
 \end{figure}

\begin{figure}
\centering
 \begin{minipage}{0.5\textwidth}
  \includegraphics[width=1\textwidth]{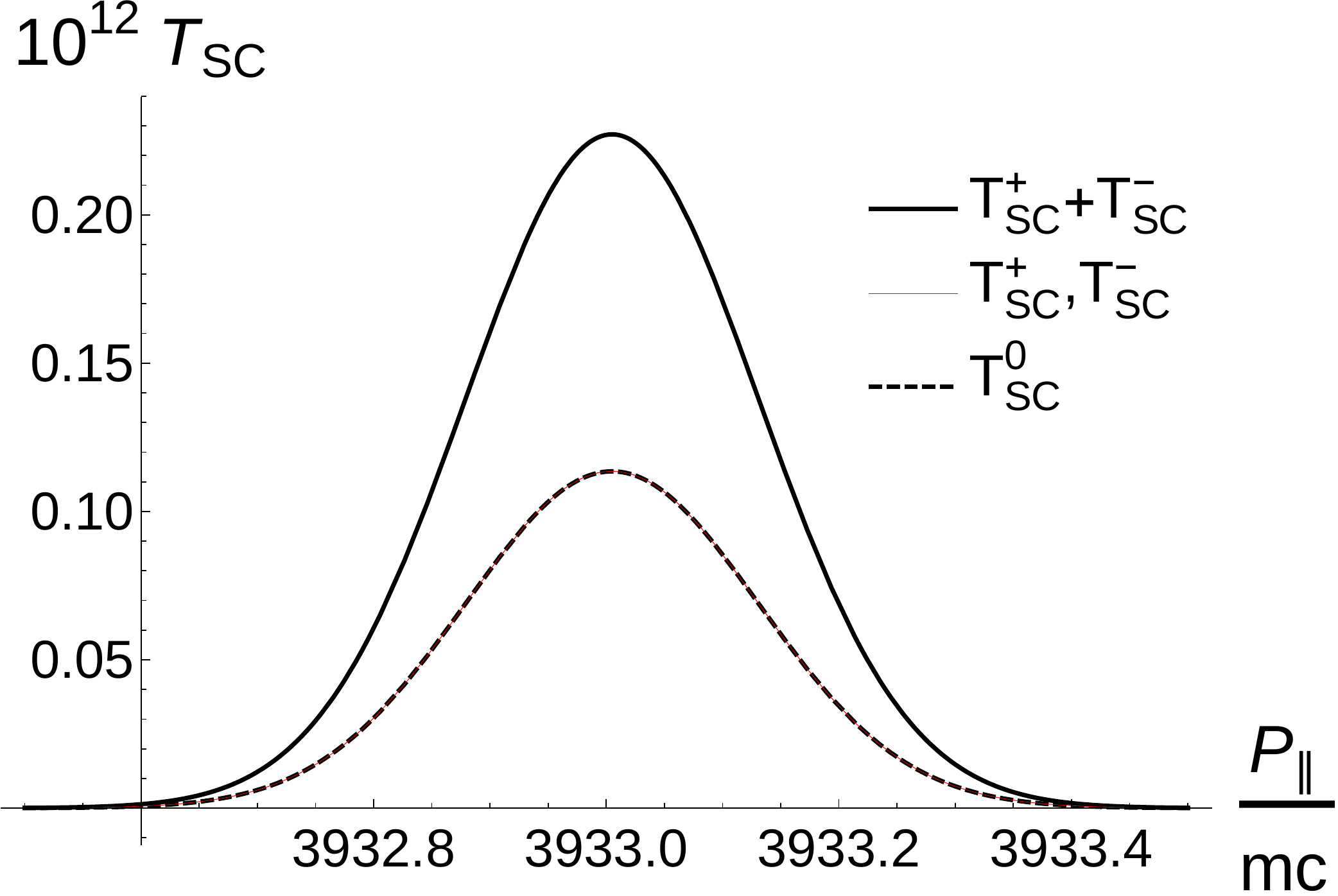} 
 \end{minipage}
\begin{minipage}{0.49\textwidth}
  \includegraphics[width=1\textwidth]{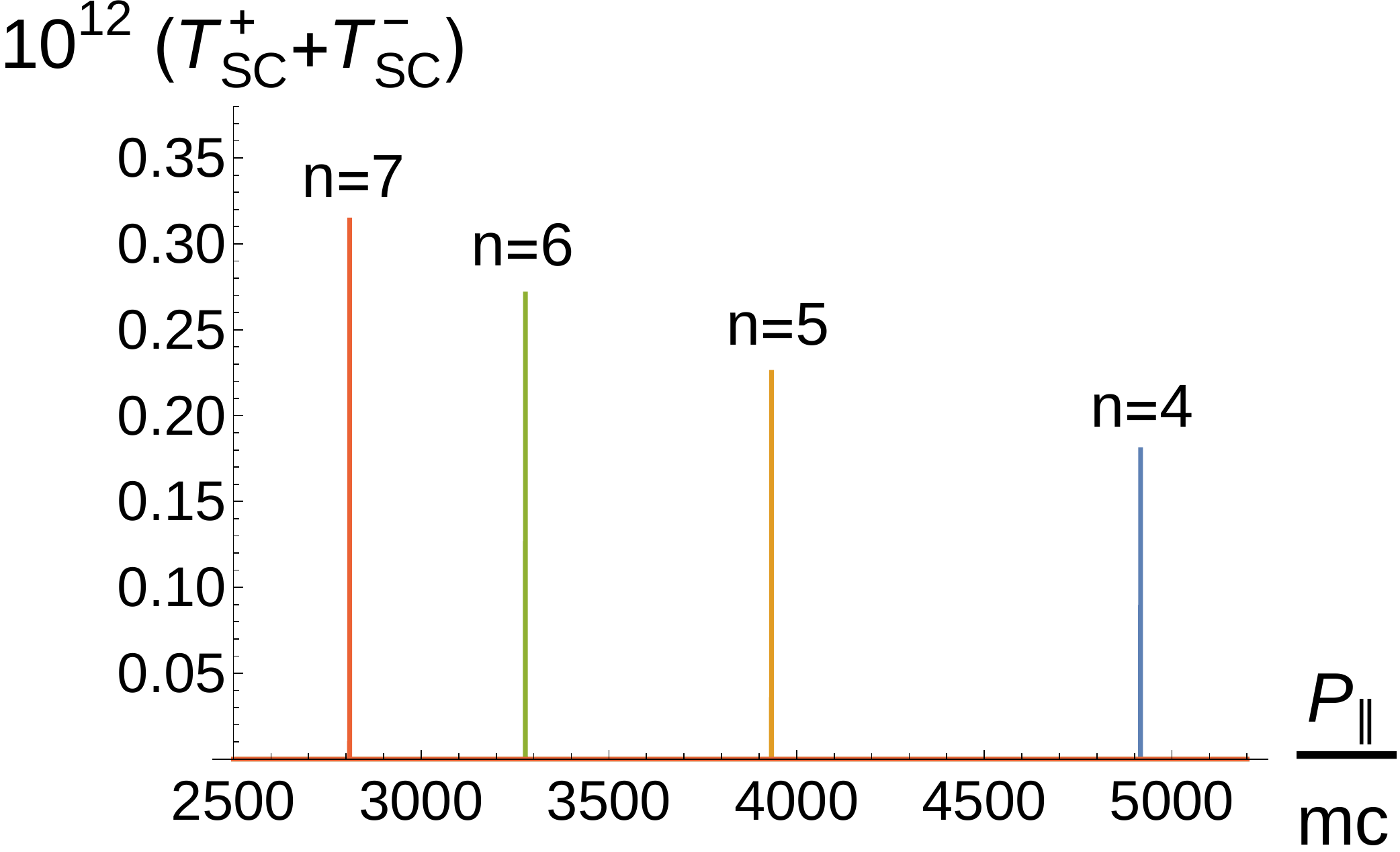}
 \end{minipage}
 \caption{ Non-interference part of the momentum spectrum of pairs produced by the rotating electric field pulse given in (\ref{eq:constrectpulse}) for an amplitude field strength \(\epsilon=0.1\), transversal momentum \(P_z=0\), a pulse length of \(\tau=10\,\text{fs}\) and for \(\sigma= 2\pi\,n\) which corresponds to a frequency of \(\nu=n\cdot100\,\text{THz}\). The left side shows the spectrum for \(n=5\) in scalar QED (dashed) as well as in spinor QED (solid). The two different spin states are plotted as a thin red line. As already seen in Fig.~\ref{fig:MS_t=infty} for this range of parameters the only difference between the spin states and the scalar case are interference effects. On the right side the spinor spectrum is plotted for \(n=4,5,6,7\). One sees that the radial momentum \(P_\parallel\) of the produced pairs depends characteristically on the laser parameters. }
  \label{fig:T}
 \end{figure}

In this Section we compute the pair creation rate for the rotating electric field which is described by
\begin{align}
 \vec{E}=E_0 (\cos(\omega t),-g\sin(\omega t),0),\label{eq:Erot}
\end{align}
where \(g=\pm1\) defines the sense of rotation.
It is possible to compute the integrals \(K_{s}(t_p)\) analytically in terms of elliptic integrals (this was shown for the scalar case in \cite{paper1}). To do so we first look at the turning points. These are given by \cite{paper1}
\begin{align}
 \omega t_k^\pm=\arcsin\left(\frac{P_x}{P_\parallel}\right)\pm\i\,\text{arcosh}\left(\frac{\left(c^2 P_\parallel^2+\epsilon_\perp^2\right)\left(\frac{\lambda_c\omega}{c\epsilon}\right)^2+m^2c^4}{2\left(\frac{\lambda_c\omega}{c\epsilon}\right)\,  cP_\parallel\, mc^2}\right)+2\pi k. \label{eq:rotatingcompltp}
\end{align}
We find an infinite number of turning points. 
They all have the same distance to the real axis and thus the integral \(K_\pm(t_p)\) defined in \Eq({\ref{eq:K_s}}) has the same imaginary part for each of this pairs of turning points and we thus define
\begin{align}
\kappa_s:=-\Im[K_s(t_k^+)]. \label{eq:kappa}
\end{align}
Since the imaginary parts of the turning points are equidistant from each other, the real parts of \(K_\pm(t_k^+)\) are multiples of
\begin{align}
\theta_s:=\Re[K_s(t_{k+1}^+)]-\Re[K_s(t_{k}^+)] \label{eq:theta}
\end{align}
for different turning points (after subtraction of a global phase). The exact form of the integrals for \(K_\pm(t_p)\) is derived in Appendix \ref{app:RotAna} and is given in \Eqs (\ref{eq:Krot})-(\ref{eq:K_xyrot}) as functions of the elliptic integrals \Eqs(\ref{eq:K})-(\ref{eq:Pi}).\\
However since there is an infinite number of turning points, which give the same contribution to the momentum spectrum, the sum in \Eq(\ref{eq:MomentumSpectrum}) diverges. This only happens because the fields rotates infinitely long which is unphysical. If we however consider a rectangular pulse, i.e. we turn the field on at time \(t_i\) and off at \(t_f\), only turning points which satisfy 
\begin{align}
t_i<\Re(t_k^\pm)<t_f \label{eq:criterium}
\end{align}
contribute to the momentum spectrum. We will therefore study the rotating rectangular pulse given by\footnote{Observe that the semiclassical solution (\ref{eq:solmulti}) of the multiple integral iteration (\ref{eq:multiint}) is only correct for analytic potentials. For potentials with discontinuities in their derivatives the semiclassical limit is dominated by the first term \cite{Berry1982}. However since we are interested in the rotation and not in the effects off turning on and off the field we will treat the potential as if it was analytical. Note however that switching on and off effects have been recently studied in the context of pair creation in an exponentially decreasing field \cite{Adorno2014}. \phantomsection \label{fn:nonanalytical}}
\begin{align}
 \vec{E}=E_0 \Rect\left(\frac{t}{\tau}\right)(\cos(\omega t),-g\sin(\omega t),0), \label{eq:constrectpulse}
\end{align}
where we defined the rectangular box function
\begin{align}
 \Rect(x)=\Theta(x)-\Theta(x-1)
\end{align}
and set \(t_i=0,\,t_f=\tau\) for convenience.\\
Using the criterion (\ref{eq:criterium}) and \Eq(\ref{eq:rotatingcompltp}) we find that for every full period of \(2\pi/\omega\) inside the pulse length \(\tau\), one turning point contributes fully to the momentum spectrum. If we now introduce  the parameter \(\sigma=\omega\tau\), the momentum spectrum (\ref{eq:MomentumSpectrum}) with \(n\) turning points taken into account, i.e. for \(\sigma= 2\pi\,n\), takes the form
\begin{align}
W^s_\text{rot}(\vec{P})=\left(n+2\sum_{i=1}^{n-1}(n-i) \cos(i \theta_s)\right)\e^{-2\kappa_s}, \label{eq:MSrectangularpulse}  
\end{align}
where \(\kappa_s\) and \(\theta_s\) are defined in \Eqs(\ref{eq:kappa}) and (\ref{eq:theta}) respectively.\\
From \Eq(\ref{eq:MSrectangularpulse}) we find that interference effects occur for \(n>1\). These effects can for example be seen in the momentum spectra in the spinor and scalar case which are plotted in Fig.~\ref{fig:MS_t=infty} for two different ranges of parameters. \\
The parameters of on the left side of Fig.~\ref{fig:MS_t=infty} are comparable to the ones chosen in \cite{Blinne2013}. In addition to the interference effects reported in that work one can also see that the spinor particle rate is increased with respect to the scalar one for particles with one spin and decreased for the others. As explained in Section \ref{sec:SCMomSpec} this is due to the factor \(\exp{(\pm\i K_{yx}(t_p))}\). Looking at the solution of \(K_{yx}(t_p)\) given in \Eq(\ref{eq:K_xyrot}) we find that there is a global factor of \(g\) which controls the sense of the rotation of the field (\ref{eq:Erot}). We thus find that the sense of the rotation determines  which spin state the spectrum is dominated by.\\
On the right side of Fig.~\ref{fig:MS_t=infty} a part of the spectrum is plotted for parameters which are closer to the ones of present-day laser systems, i.e.~a pulse length \(\tau=10\,\text{fs}\) and a laser frequency \(\nu=2\pi\, \omega=n\cdot100\,\text{THz}\) while the amplitude field strength is still \(\epsilon=0.1\). Although we still find interference effects the pattern is to narrow to be resolved by a detector. We therefore look at the non-interference part of the spectrum
 \begin{align}
  T^s_\text{rot}(\vec{P})=n\,\e^{-2\kappa_s},  
 \end{align}
 which is plotted in Fig.~\ref{fig:T}. One sees that the difference between the spin states lies only in the interference effects for this range of parameters. Accordingly the non-interference part of the spectrum is the same for either spin state and the scalar case. We also find that the radial momentum \(P_\parallel\) is tightly confined by the laser frequency \(\nu\).
 
\section{Total pair creation of the constant rotating field}
\label{sec:TotalParticleYield}
\begin{figure}
 \begin{minipage}{0.8\textwidth}
  \includegraphics[width=\textwidth]{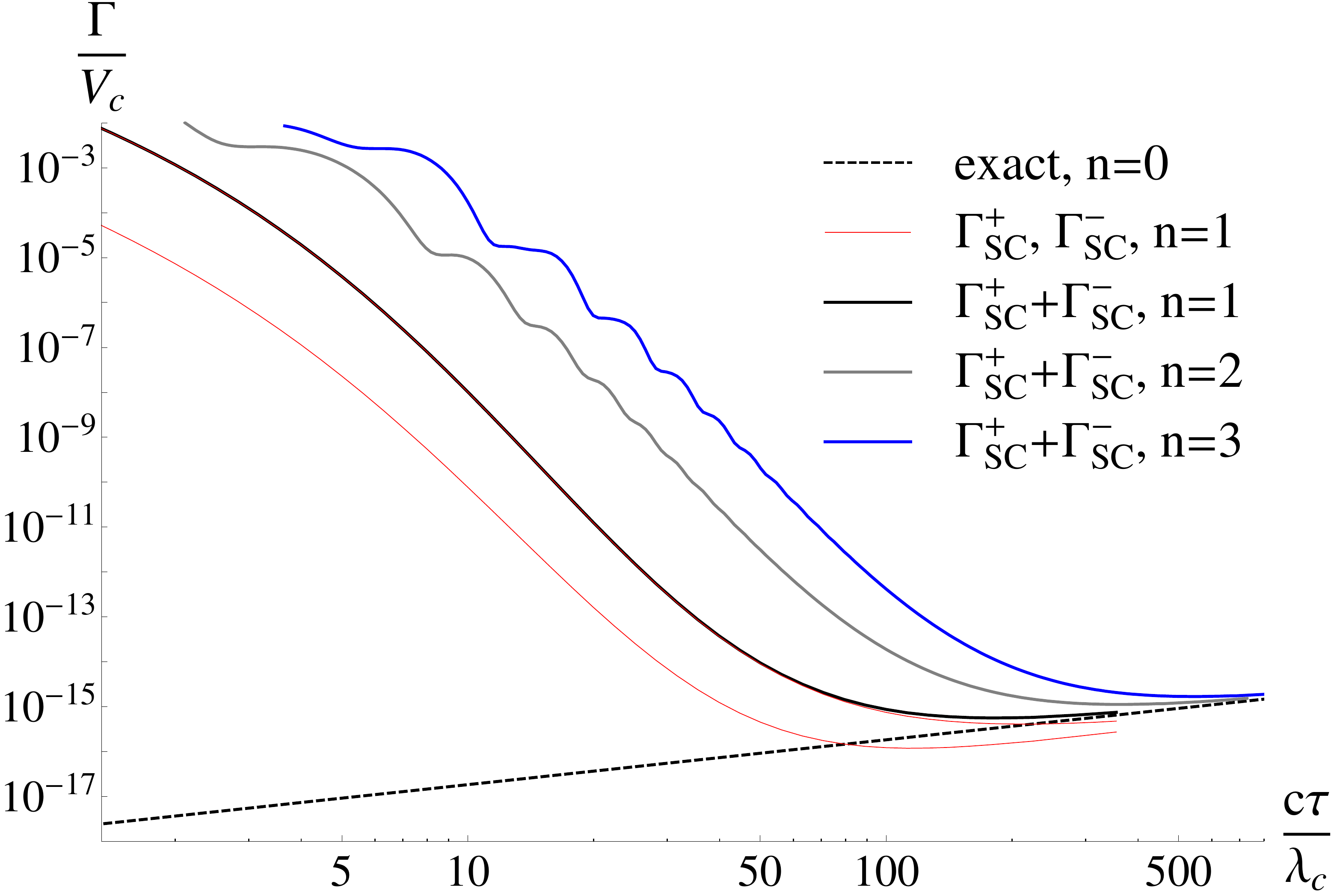}
 \end{minipage}
  \caption{Total particle number per Compton volume \(V_c\) of the rotating rectangle pulse for \(\epsilon=0.1\) and \(\sigma=2\pi\,n,\,n=0,1,2,3\) as a function of the pulse length \(\tau\). For \(n=1\) we also plotted the pair creation in the respective spin states. One sees that the pair creation is dominated by one of them for a wide range of parameters since the line for one spin direction overlaps the sum of both up to \(\tau\sim60\). Observe that there are interference effects for \(n>1\).} 
  \label{fig:TotalYield}
 \end{figure}

The total particle production per volume can be calculated by integrating over the momentum spectrum
\begin{align}
 \frac{\Gamma^s}{V}=\int  \frac{d^3P}{(2\pi \hbar )^3}\, W^s_\text{SC}\left(\vec{P}\right).
\end{align}
This integration normally has to be performed numerically.\\
For the rotating rectangle pulse we study parameters for which \(\sigma= 2\pi\,n\). As discussed in Section \ref{sec:rotating} for this cases \(n\) pairs of turning points contribute to the pair creation rate. In Fig.~\ref{fig:TotalYield} the total particle yield for spinor QED is plotted for \(n=1,2,3\). For \(n=1\) we also show how the total yield is distributed with respect to spin. As already mentioned in \ref{sec:rotating} one finds that the pair creation rate is dominated by pairs with one spin depending on the sense of rotation of the field \(g\) for a wide range of pulse lengths \(\tau\). For \(\tau \rightarrow \infty\) the ratio of spin states goes to \(1:1\). This is due to the fact that for longer pulses the effect of the rotation is smaller. Accordingly for \(\tau \rightarrow \infty\) the results are asymptotic to those of the non-rotating rectangle pulse with length \(\tau\) which has a total pair creation rate of \footnote{Here we again ignore discontinuities 
in the derivative of the potential, see footnote \ref{fn:nonanalytical} on page \pageref{fn:nonanalytical}.}
\begin{align}
 \Gamma=2 \frac{V}{V_c} \frac{\tau \epsilon^2}{(2\pi)^3} \e^{-\frac{\pi}{\epsilon}},
\end{align}
where \(V_c=\lambda_c^3\) is the Compton volume. Since the non-rotating rectangle pulse is an one-component field, half of the particles have either spin. We find that for \(n=0\) there are no interference effects in the total particle number. This is clear since only one turning point contributes. As in the momentum spectrum we see that there are interference effects in the total particle number  for \(n>1\).\\
\section{Conclusions and Remarks}
\label{sec:conclusions}
We introduce a semiclassical treatment of the Dirac-equation for two-component time-dependent electric fields. In the limit of one-component fields our result reduces to the well known WKB-approximation. We use this method to investigate the momentum spectrum and total pair creation of Schwinger pair production semiclassically.\\
We find that in difference to the one-component case, the pair creation rate for spinor QED can be dominated by pairs of one spin direction. Due to this fact the pair creation rate of spinor QED differs from the one of scalar QED (treated in \cite{paper1} for two-component fields) even for the case of one (dominant) pair of turning points. This is evident in rotating field pulses, where the spin state that most pairs are created in, depends on the sense of rotation of the field for a certain range of pulse lengths and frequencies.\\
For a constant rotating field the momentum spectrum can be computed analytically. However one has to use a rectangle pulse to get a finite pair creation rate. If one chooses the parameter \(\sigma= 2\pi\,n\) the momentum spectrum turns out to be rotationally invariant in the \(P_x\text{-}P_y\) plane  as was also found for the Sauter pulse for high enough \(\sigma\) \cite{Blinne2013}. This can be explained by the fact that for long enough pulses the effects of the pulse shape get less important and the effects of rotation are more apparent. As for the Sauter pulse we find interference effects in the momentum spectrum of the produced pairs. These effects become stronger for higher \(\sigma\). In addition to this we also find interference effects in the total particle number. \\
As was discussed in \cite{Blinne2013}, the characteristic momentum spectra for rotating pulses could be a decisive fingerprint for Schwinger pair creation if the QED cascade seeded by a created pair can be distinguished from one seeded by vacuum impurities. We showed that additionally to the form of the momentum spectrum, pairs have a preferred spin direction in the parameter range chosen in \cite{Blinne2013}. For pulse lengths and frequencies reached by laser systems at the moment, both the characteristic interference patterns and the spin dependence could not be resolved by a detector. However the radial momentum spectrum of created pairs is very narrow. Depending on the parameters of future laser systems these results can also help with the differentiation discussed above.
\section*{Acknowledgements} 
The authors thank Antonino Di Piazza, Christoph H. Keitel, Hagen Kleinert and Clément Stahl for fruitful discussions. 
ES is supported by the Erasmus Mundus Joint Doctorate Program by Grant Number 2012-1710 from the EACEA of the European Commission.

\appendix

\section{Ansatz for the Dirac equation}
\label{app:Ansatz}
The ansatz (\ref{eq:ansatz1})-(\ref{eq:ansatzdot2}) can be derived from the usual treatment for the one-component case (see e.g. \cite{Dumlu2011}). There one uses the WKB-solution and introduces the coefficients \(\alpha(t)\) and \(\beta(t)\) for the positive and negative solution of the zeroth order in \(\hbar\) respectively. Putting this in the Dirac equation one finds coupled differential equations where the time derivative of \(\alpha(t)\) depends only on \(\beta(t)\) and vice versa.\\
Since the zeroth order WKB solution in the two-component case is equivalent to the one-component case, we can make the ansatz
\begin{align}
 \psi^{\pm}_{i}(t)&=\frac{\alpha^\pm(t)}{\sqrt{\scE(t)}}F_{\alpha i}^\pm(t) \e^{-\i K(t)}+\frac{\beta^\pm(t)}{\sqrt{\scE(t)}}F_{\beta i}^\pm(t) \e^{\i K(t)}, \label{eq:firstansatz1}\\
 \dot{\psi}^{\pm}_{i}(t)&=-\i\frac{\scE(t)}{\hbar}\left(\frac{\alpha^\pm(t)}{\sqrt{\scE(t)}}F_{\alpha i}^\pm(t) \e^{-\i K(t)}-\frac{\beta^\pm(t)}{\sqrt{\scE(t)}}F_{\beta i}^\pm(t) \e^{\i K(t)}\right)  \label{eq:firstansatz2},
\end{align}
where we introduced the coefficients \(F_{\nicefrac{\alpha}{\beta}i}^\pm(t)\) where \(i=1,2\) and used \(K(t)\) defined in \Eq(\ref{eq:Kint}). We now want to choose these coefficients such that the differential equations for \(\alpha^\pm(t)\) and \(\beta^\pm(t)\) take the same form as in the one-component case. To do so we use (\ref{eq:firstansatz1}) and (\ref{eq:firstansatz2}) in the coupled Dirac equations (\ref{eq:coupledDirac1}) and (\ref{eq:coupledDirac2}) to find
\begin{align}
 \dot{\alpha}^\pm(t) F_{\alpha1}^\pm(t)=\left(\dot{\scE}(t)F_{\beta1}^\pm(t)\mp\cv \dot{p}_{x-y}(t)F_{\beta2}^\pm(t)\right)\frac{\beta^\pm(t)}{2\scE(t)}\e^{2\i K(t)}-\left(2\scE(t)\dot{F}_{\alpha1}^\pm(t)\pm\cv \dot{p}_{x-y}(t)F_{\alpha2}^\pm(t)\right)\frac{\alpha^\pm(t)}{2\scE(t)},  \label{eq:ansatzindirac1} \\
 \dot{\alpha}^\pm(t) F_{\alpha2}^\pm(t)=\left(\dot{\scE}(t)F_{\beta2}^\pm(t)\mp\cv \dot{p}_{x+y}(t)F_{\beta1}^\pm(t)\right)\frac{\beta^\pm(t)}{2\scE(t)}\e^{2\i K(t)}-\left(2\scE(t)\dot{F}_{\alpha2}^\pm(t)\pm\cv \dot{p}_{x+y}(t)F_{\alpha1}^\pm(t)\right)\frac{\alpha^\pm(t)}{2\scE(t)}. \label{eq:ansatzindirac2}
\end{align}
As discussed before we now want to choose \(F_{\nicefrac{\alpha}{\beta}i}^\pm(t)\) such that the last term in (\ref{eq:ansatzindirac1}) and (\ref{eq:ansatzindirac2}) vanishes. This leads to
\begin{align}
 \dot{F}_{\alpha1}^\pm(t)=\mp\frac{\cv \dot{p}_{x-y}(t)}{2\scE(t)}F_{\alpha2}^\pm(t), &&
 \dot{F}_{\alpha2}^\pm(t)=\mp\frac{\cv \dot{p}_{x+y}(t)}{2\scE(t)}F_{\alpha1}^\pm(t). \label{eq:Fa1steq}
\end{align}
Additionally \Eqs (\ref{eq:ansatzindirac1}) and (\ref{eq:ansatzindirac2}) should not contradict each other and thus
\begin{align}
\frac{\dot{\scE}(t)F_{\beta1}^\pm(t)\mp \cv \dot{p}_{x-y}(t)F_{\beta2}^\pm(t)}{F_{\alpha1}^\pm(t)}=\frac{\dot{\scE}(t)F_{\beta2}^\pm(t)\mp \cv \dot{p}_{x+y}(t)F_{\beta1}^\pm(t)}{F_{\alpha2}^\pm(t)}. \label{eq:Fb2ndeq}
\end{align}
By analogy we find 
\begin{align}
 \dot{F}_{\beta1}^\pm(t)=\pm\frac{\cv \dot{p}_{x-y}(t)}{2\scE(t)}F_{\beta2}^\pm(t), &&
 \dot{F}_{\beta2}^\pm(t)=\pm\frac{\cv \dot{p}_{x+y}(t)}{2\scE(t)}F_{\beta1}^\pm(t). \label{eq:Fb1steq}
\end{align}
and
\begin{align}
 \frac{\dot{\scE}(t)F_{\alpha1}^\pm(t)\pm \cv \dot{p}_{x-y}(t)F_{\alpha2}^\pm(t)}{F_{\beta1}^\pm(t)}=\frac{\dot{\scE}(t)F_{\alpha2}^\pm(t)\pm \cv \dot{p}_{x+y}(t)F_{\alpha1}^\pm(t)}{F_{\beta2}^\pm(t)}. \label{eq:Fa2ndeq}
\end{align}
Using 
\begin{align}
 \dot{\scE}(t)\scE(t)=\cv \dot{p}_x(t) \cv p_x(t) + \cv \dot{p}_y(t) \cv p_y(t)
\end{align}
in \Eq(\ref{eq:Fb2ndeq}), sorting by terms proportional to the derivative of the momenta and requiring \(\dot{p}_x(t)\) and \(\dot{p}_y(t)\) to be independent we find two equations for \(F_{\alpha1}^\pm(t)/F_{\alpha2}^\pm(t)\). Setting these equal leads to
\begin{align}
 \frac{\cv p_{x+y}(t)}{\scE(t)}\left(F_{\beta1}^\pm(t)\right)^2\pm2F_{\beta1}^\pm(t)F_{\beta2}^\pm(t)+\frac{\cv p_{x-y}(t)}{\scE(t)}\left(F_{\beta1}^\pm(t)\right)^2=0.
\end{align}
A solution of this equation is given by
\begin{align}
 \frac{F_{\beta1}^\pm(t)}{F_{\beta2}^\pm(t)}=\frac{\pm\scE(t)\varpm\epsilon_\perp}{\cv p_{x+y}(t)}=\frac{\cv p_{x-y}(t)}{\pm\scE(t)\varmp\epsilon_\perp}. \label{eq:Fb1Fb2}
\end{align}
Using this and 
\begin{align}
 2\dot{\scE}(t)\scE(t)=\cv \dot{p}_{x+y}(t) \cv p_{x-y}(t) + \cv \dot{p}_{x-y}(t) \cv p_{x+y}(t)
\end{align}
in \Eq(\ref{eq:Fb2ndeq}) we find
\begin{align}
 \frac{F_{\alpha1}^\pm(t)}{F_{\alpha2}^\pm(t)}=\frac{\mp\scE(t)\varpm\epsilon_\perp}{\cv p_{x+y}(t)}=\frac{\cv p_{x-y}(t)}{\mp\scE(t)\varmp\epsilon_\perp}. \label{eq:Fa1Fa2}
\end{align}
One can check that the solutions (\ref{eq:Fb1Fb2}) and (\ref{eq:Fa1Fa2}) also fulfill \Eq(\ref{eq:Fa2ndeq}). This means we are left with the three equations combining \({F_{\alpha1}^\pm(t)}\) and \({F_{\alpha2}^\pm(t)}\) as well as \({F_{\beta1}^\pm(t)}\) and \({F_{\beta2}^\pm(t)}\) respectively.\\
If we now use \Eq(\ref{eq:Fa1Fa2}) in the first equation of (\ref{eq:Fa1steq}) we find
\begin{align}
  \dot{F}_{\alpha1}^\pm(t)&=\frac{1}{2}\frac{\dot{p}_{x-y}(t)}{p_{x-y}(t)}\frac{\scE(t)\pmmp\epsilon_\perp}{\scE(t)}F_{\alpha1}^\pm(t)\\
  &
  =\frac{1}{2}\left[\frac{\dot{p}_{x-y}(t)}{p_{x-y}(t)}
  \pmmp\epsilon_\perp\left(\frac{\dot{\scE}(t)}{\scE(t)^2-\epsilon_\perp^2}+\i \frac{\dot{p}_x(t)p_y(t)-\dot{p}_y(t)p_x(t)}{\scE(t)p_\parallel(t)^2}\right)\right]F_{\alpha1}^\pm(t).
\end{align}
We can integrate this equation to find
\begin{align}
 F_{\alpha1}^\pm(t)=C_{\alpha}^\pm\frac{\sqrt{cp_{x-y}(t)}}{\sqrt{\scE(t)\pmmp\epsilon_\perp}}\sqrt{cp_{\parallel}(t)}\exp\left(\pmmp \frac{\i}{2}K_{xy}(t)\right), \label{eq:Fa1sol}
\end{align}
where the integral \(K_{xy}(t)\) is defined in \Eq(\ref{eq:K_xy}).\\
Using \Eq(\ref{eq:Fa1Fa2}) again we find
\begin{align}
 F_{\alpha2}^\pm(t)=\mp C_{\alpha}^\pm\frac{\sqrt{cp_{x+y}(t)}}{\sqrt{\scE(t)\mppm\epsilon_\perp}}\sqrt{cp_{\parallel}(t)} \exp\left(\pmmp \frac{\i}{2}K_{xy}(t)\right). \label{eq:Fa2sol}
\end{align}
One can now check if the two solutions (\ref{eq:Fa1sol}) and (\ref{eq:Fa2sol}) fulfill the remaining equation namely the second equation of (\ref{eq:Fa1steq}). This is the case and thus we can use them in our ansatz. By analogy we find
\begin{align}
 F_{\beta1}^\pm(t)&=C_{\beta}^\pm\frac{\sqrt{cp_{x-y}(t)}}{\sqrt{\scE(t)\mppm\epsilon_\perp}}\sqrt{cp_{\parallel}(t)} \exp\left(\mppm \frac{\i}{2}K_{xy}(t)\right), \label{eq:Fb1sol}\\
 F_{\beta2}^\pm(t)&=\pm C_{\beta}^\pm\frac{\sqrt{cp_{x+y}(t)}}{\sqrt{\scE(t)\pmmp\epsilon_\perp}}\sqrt{cp_{\parallel}(t)} \exp\left(\mppm \frac{\i}{2}K_{xy}(t)\right). \label{eq:Fb2sol}
\end{align}
We can now put these results into \Eq(\ref{eq:ansatzindirac1}) to find 
\begin{align}
 \dot{\alpha}^\pm(t)&=\frac{\pmmp\dot{\scE}(t)\epsilon_\perp-\i c^2\left[\dot{p}_x(t)p_y(t)-\dot{p}_y(t)p_x(t)\right]}{2 cp_\parallel(t)\scE(t)}\frac{C_\beta^\pm}{C_\alpha^\pm}\e^{ \i K_{\mppm}(t)}\beta^\pm(t). \label{eq:alphaC}
\end{align}
To get the value of \(C_\alpha^\pm/C_\beta^\pm\) we can compare to the one-component case of \cite{Dumlu2011}. To do so we use \(p_y(t)=P_y\) to find 
\begin{align}
 \left.e^{\pm \i K_{xy}(t)}\right|_{p_y(t)=P_y}=\frac{\scE(t)cP_y\pm\i\epsilon_\perp cp_x(t)}{\sqrt{(\scE(t)cP_y)^2+(\epsilon_\perp c p_x(t))^2}}, \label{eq:comptoonecomp}
\end{align}
Using this in \Eq(\ref{eq:alphaC}) we find
\begin{align}
  \left.\dot{\alpha}^\pm(t)\right|_{p_y(t)=P_y}&=-\i\frac{c\dot{p}_x(t)}{2\scE(t)^2}\sqrt{\epsilon_\perp^2+cP_y^2}\frac{C_\beta^\pm}{C_\alpha^\pm}\e^{\i K(t)}\beta^\pm(t)
\end{align}
This result is equivalent to \Eq(14) of \cite{Dumlu2011}   if \(C_\alpha^\pm=-\i C_\beta^\pm:=C^\pm\). Using this result in \Eqs(\ref{eq:firstansatz1}) and (\ref{eq:firstansatz2}) leads to the ansatz used in section \ref{sec:SCMomSpec}, i.e. \Eqs(\ref{eq:ansatz1})-(\ref{eq:ansatzdot2}).\\

\section{Analytic calculation of the momentum spectrum}
\label{app:ana}
\label{app:RotAna}
In this Appendix we compute the integrals \(K(t_k)\) and \(K_{xy}(t_k)\) given by \Eqs(\ref{eq:Kint}) and (\ref{eq:K_xy}) for the constant rotating field (\ref{eq:Erot}). Note that the imaginary part of \(K(t_k)\) was already computed in \cite{paper1}. For this purpose we introduce the adiabatic parameter
 \begin{align}
 \gamma:=\frac{\omega m c}{ e E_0}=\frac{\lambda_c\omega}{c}\frac{E_c}{E_0}.
 \end{align}\\  
For the calculation of \(K(t_k)\) we first change the variable of the integral to the dimensionless \(T=\omega t \) such that \Eq(\ref{eq:Kint}) takes the form
\begin{align}
K(t):=\frac{2}{\hbar\omega}\int_{0}^{\omega t} \scE(T') dT'. \label{eq:Kint2}
\end{align}
We now change the variable to 
\begin{align}
 x=-\frac{cp_x(T)^2+cp_y(T)^2}{\epsilon_\perp^2}, \label{eq:sub}
\end{align}
where for the rotating electric field defined in \Eq(\ref{eq:Erot}) the kinetical momenta are given by
\begin{align}
 p_x(T)=P_x-\frac{mc}{\gamma}\sin(T), && p_y(T)=P_y-g\frac{mc}{\gamma}\cos(T).
\end{align}
For this special field configuration we find (for details see \cite{paper1})
\begin{align}
 \frac{d x}{d T}=\pm \i \sqrt{(x-y_-^2)(x-y_+^2)}, \label{eq:dxdT}
\end{align}
where we defined
\begin{align}
y_{\pm}:=\i\frac{\gamma cP_\parallel\pm mc^2}{\gamma \epsilon_\perp},&&
P_\parallel:=\sqrt{P_x^2+P_y^2}.
\end{align}
One finds that the sign in \Eq(\ref{eq:dxdT}) changes at \(x=y_-^2=y_+^2\) and that the new variable takes the values \(x(\Re(t_k))=y_-^2\) and \(x(t_k)=1\). Thus performing the substitution (\ref{eq:sub}) the integral (\ref{eq:Kint2}) takes the form
\begin{align}
 K(t_k)=\i\frac{2\epsilon_\perp}{\hbar\omega}\left(\int_{y_-^2}^1-2k\int_{y_-^2}^{y_+^2}\right)\frac{\sqrt{1-x}\,dx}{\sqrt{(x-y_-^2)(x-y_+^2)}}+\Phi , \label{eq:KwithInts}
 \end{align}
where \(\Phi\) is a global phase factor which stems from integrating from \(x(0)\) to the turning point \(x(t_0)\). The first integral corresponds to the integral parallel to the imaginary \(t\)-axis and will give rise to the imaginary part of \(K(t_k)\) while the second one gives the real contribution. \\
It is possible to solve the integrals in \Eq(\ref{eq:KwithInts}) leading to (see \cite{Grad2000} Eqs.~3.141.2 and 3.141.5)
\begin{align}
 K(t_k)&=\i\frac{4\epsilon_\perp}{\hbar\omega}\sqrt{1-y_+^2}\left[\E\left(\sqrt{\frac{1-y_-^2}{1-y_+^2}}\right)-\K\left(\sqrt{\frac{1-y_-^2}{1-y_+^2}}\right)-2k\,\i \E\left(\sqrt{\frac{y_-^2-y_+^2}{1-y_+^2}}\right)\right]+\Phi, \label{eq:Krot}
\end{align}
were we use the elliptic integrals (see \cite{Grad2000} Eqs.~8.111-112)
\begin{align}
 \K(k):=&\int_0^{\pi/2}\frac{d \theta}{\sqrt{1-k^2\sin^2(\theta)}},\label{eq:K}\\
 \E(k):=&\int_0^{\pi/2}{\sqrt{1-k^2\sin^2(\theta)}}{d \theta},\label{eq:E}\\
 \PiI(n,k):=&\int_0^{\pi/2}\frac{d \theta}{\left(1-n\sin^2(\theta)\right)\sqrt{1-k^2\sin^2(\theta)}}.\label{eq:Pi}
 \end{align}
We can compute the integral \(K_{xy}(t_k)\) in an analogous way. From \Eq(\ref{eq:K_xy}) follows
 \begin{align}
K_{xy}(t)&:=\epsilon_\perp\int_{0}^{\omega T} \frac{p_x'(T')p_y(T')-p_y'(T')p_x(T')}{\scE(T')p_\parallel(T')^2}dT' \label{eq:K_xy2}.
 \end{align}
This can be brought into the form
\begin{align}
 K_{xy}(t_k)=\i\frac{g}{2}\left(\int_{y_-^2}^1-2k\int_{y_-^2}^{y_+^2}\right)\frac{\left(x-y_- y_+\right)\,dx}{x\sqrt{(x-y_-^2)(x-y_+^2)}\sqrt{1-x}}+\Phi_{xy},
\end{align}
Where \(\Phi_{xy}\) is again a global phase. So that we find (see \cite{Grad2000} Eqs.~3.131.3, 3.131.6, 3.137.4 and 3.137.6)
\begin{align}
\begin{split}
 K_{xy}(t_k)=&-\frac{\i g}{\sqrt{1-y_+^2}}
 \left[\K\left(\sqrt{\frac{1-y_-^2}{1-y_+^2}}\right)-y_-y_+\PiI\left(1-y_-^2,\sqrt{\frac{1-y_-^2}{1-y_+^2}}\right)\right]+\Phi_{xy}\\
 &+2k\frac{\,g}{\sqrt{1-y_+^2}}\left[\left(1-y_-y_+\right)\K\left(\sqrt{\frac{y_-^2-y_+^2}{1-y_+^2}}\right)+({y_-^2-1})\frac{y_+}{y_-}\PiI\left(\frac{1}{y_-^2}\frac{y_-^2-y_+^2}{1-y_+^2},\sqrt{\frac{y_-^2-y_+^2}{1-y_+^2}}\right)\right]. \label{eq:K_xyrot}
 \end{split}
\end{align}


\end{document}